\documentclass[aps,pra,reprint,superscriptaddress,amsmath,amssymb,preprintnumbers,showpacs]{revtex4-1}
\usepackage[normalem]{ulem} 

\usepackage{enumerate}
\usepackage{graphicx}
\usepackage{xcolor}
\usepackage{float} 
\usepackage{xspace}
\usepackage{bm} 
\graphicspath{{./}}

\newcommand{\ket}[1]{\bigl| #1 \bigr>} 
\newcommand{\bra}[1]{\bigl< #1 \bigr|} 
\newcommand{\braket}[2]{\bigl< #1 \vphantom{#2} \bigr|
 \bigl. #2 \vphantom{#1} \bigr>} 
\newcommand{\abs}[1]{\left| #1 \right|} 

\newcommand{\floqel}[1]{\xi_{E_\text{e}}}
\newcommand{\floqnuc}[1]{\xi_{E_\text{N}}}

\renewcommand{\Im}{\text{Im}}
\newcommand{\op}[1]{\hat{#1}}
\newcommand{\vop}[1]{\hat{\vec{#1}}}
\renewcommand{\vec}[1]{{\mathbf{\bm{#1}}}}
\newcommand{\uv}[1]{\hat{\mathbf{#1}}}
\newcommand{\hBN}{h-BN}
\newcommand{\berryc}{\mathcal{A}}
\newcommand{\LG}{\text{LG}}
\newcommand{\VG}{\text{VG}}

\begin{document}
\title{Structure- and laser-gauges for the semiconductor Bloch equations in high-harmonic generation in solids} 
\author{Lun \surname{Yue}}
\email{lun\_yue@msn.com}
\author{Mette B. \surname{Gaarde}}
\email{mgaarde1@lsu.edu}
\affiliation{Department of Physics and Astronomy, Louisiana State University, Baton Rouge, Louisiana 70803-4001, USA}
\date{\today}

\begin{abstract}
  The semiconductor Bloch equations (SBEs) are routinely used for simulations of strong-field laser-matter interactions in condensed matter. In systems without inversion or time-reversal symmetries, the Berry connections and transition dipole phases (TDPs) must be included in the SBEs, which in turn requires the construction of a smooth and periodic structure gauge for the Bloch states. Here, we illustrate a general approach for such a structure-gauge construction for topologically trivial systems. Furthermore, we investigate the SBEs in the length and velocity gauges, and discuss their respective advantages and shortcomings for the high-harmonic generation (HHG) process. We find that in cases where we require dephasing or separation of the currents into interband and intraband contributions, the length gauge SBEs are computationally more efficient. In calculations without dephasing and where only the total current is needed, the velocity gauge SBEs are structure-gauge independent and are computationally more efficient. We employ two systems as numerical examples to highlight our findings: an 1D model of ZnO and the 2D monolayer hexagonal boron nitride (h-BN). The omittance of Berry connections or TDPs in the SBEs for h-BN results in nonphysical HHG spectra. The structure- and laser-gauge considerations in the current work are not restricted to the HHG process, and are applicable to all strong-field matter simulations with SBEs.
\end{abstract}

\maketitle

\section{Introduction} \label{sec:intro}

In the past decade, high-harmonic generation in solids irradiated by strong laser pulses has attracted considerable attention \cite{Ghimire2011, Vampa2015Nat, You2016, Ndabashimiye2016, Garg2016, Wang2017}. Solid-state harmonics provide a potential avenue towards new attosecond light-source technologies and the engineering of compact attosecond light sources \cite{Luu2015, Han2016, Vampa2017, Sivis2017, Garg2018, Yang2019}, as well as a probe of fundamental properties in a solid such as the band structure \cite{Vampa2015b}, the Berry curvatures \cite{Liu2017, Luu2018} and topological phase transitions \cite{Bauer2018, Silva2019, Chacon2019}. Several different theoretical simulation frameworks have been applied to gain insight into the HHG process in solids, including the time-dependent density functional theory \cite{Runge84, Tancogne-Dejean2017b, Hansen2017, Yu2019}, the time-dependent Schr\"{o}dinger equation \cite{Wu2015, Zhong2016, Ikemachi2017, Jin2018} and the semiconductor Bloch equations (SBEs) \cite{Golde2008, Kira2012}. The SBEs are attractive because they provide a many-body framework that allows the inclusion of excitonic effects \cite{Langer2016, Garg2016}, treatment of dephasing and population decay \cite{Kira2012, Vampa2014}, as well as a natural separation of the total current into an intraband contribution due to electron and hole motion in their respective bands and an interband contribution due to the coupling between the bands.

Several challenges remain in the efficient and accurate simulation of the HHG process using the SBEs. First, in the absence of external fields, the generally complex-valued cell-periodic functions $u_n^{\vec{k}}(\vec{r})$ from Bloch's theorem can be subject to an arbitrary gauge transformation,
\begin{equation}
  \label{eq:intro_1}
  \ket{\tilde{u}_n^{\vec{k}}} = e^{i\varphi_n^{\vec{k}}}\ket{u_n^{\vec{k}}}, \quad (\varphi_n^{\vec{k}} \text{ real})
\end{equation}
without changing the physical properties of the system. Since the $u_n^{\vec{k}}(\vec{r})$ normally are obtained from some diagonalization procedure performed separately for each crystal momentum $\vec{k}$ in the Brillouin zone (BZ), the phases $\varphi_n^{\vec{k}}$ in Eq.~\eqref{eq:intro_1} are random-valued and generally discontinuous and nonperiodic over the BZ boundaries.
Quantities required by the SBEs as input, like the generally complex-valued transition dipole or momentum matrix elements, will be affected by such a particular structure gauge choice. Quantities involving $\vec{k}$-derivatives such as the Berry connections and the Berry curvatures require a smooth and periodic gauge. Due to the complexities involved in the construction of such a structure gauge,
numerous crude approximations are often made in the SBE calculations where e.g. the transition dipole phases (TDPs) and Berry connections are omitted, often without explicitly stating that an approximation has been made (see e.g. \cite{Vampa2014, Vampa2015, Li2019, Jiang2017, Jiang2018}). Since the Berry connections and Berry curvatures are generally nonzero in systems without parity or time-reversal symmetry \cite{Xiao2010, Vanderbilt2018}, the repercussions of such approximations still remain unexplored. Recently, however, there has been increased interest in the construction of periodic structure gauges in the HHG community. For example, Jiang et al. \cite{Jiang2017, Jiang2018} realized the importance of the TDPs on the generation of even-order high harmonics in gapped graphene and ZnO. 
The structure-gauge-invariance of the SBEs were discussed in Ref.~\cite{Li2019}, with a periodic-gauge construction shown for a one-dimensional model system. The Wannier gauge \cite{Marzari2012}, where maximally localized Wannier functions are used as a basis, has also been discussed recently in the context of simulating HHG in solids with SBEs \cite{Silva2019b}. The explicit construction of a periodic gauge for a $D$-dimensional topologically trivial solid and its implications for HHG simulations using SBEs, has not yet been extensively discussed in the context of HHG in solids.

In addition to the structure-gauge considerations, the choice of the laser-gauge warrants both practical and numerical considerations. In a truncated basis involving a finite number of bands, the nonlinear response of semiconductors is known to depend on the laser-gauge choice, with the length gauge (LG) pertaining numerous advantages over the velocity gauge (VG) \cite{Aversa1995, Kuljit2007, Taghizadeh2017}: a smaller number of bands can be used in the LG compared to the VG; no unphysical divergences arise in the LG in the direct-current (DC) limit, in contrast to the VG; the dephasing and separation of the total current into an intraband and an interband contribution is naturally treated in the LG. Earlier studies were mostly concentrated on the low-order nonlinearities, e.g. first- to third-order responses \cite{Aversa1995, Taghizadeh2017, Ventura2017}, and since the underlying physics of HHG in semiconductors above the band gap energy is radically different from the low-order nonlinearities, the convergence properties of the high-harmonics in different laser gauges still remains to be investigated. In addition, since the laser-matter coupling in the LG involves the position operator $\op{\vec{r}}$, which can be treated in reciprocal space by rewritting it into terms involving $\nabla_{\vec{k}}$, a contruction of a smooth and periodic structure gauge is required.

In this work, we investigate how structure-gauge and laser-gauge choices affect the HHG process in solids simulated with SBEs in a truncated basis. For this purpose, we give an explicit procedure on how to construct a smooth and periodic structure gauge for a $D$-dimensional topologically trivial solid, and show how the Berry connections and TDPs are naturally calculated in such a gauge. We discuss how to correctly include dephasing and separation of the total current into intraband and interband parts in the two laser gauges. We find that in terms of computational complexity, if dephasing and separation of current are desirable, the LG SBEs scale favourably compared to the VG SBEs; while if dephasing and separation of current is not needed, the LG SBEs scale unfavourably compared to the VG SBEs. We test our computational schemes on two examples: a 1D Mathieu model system mimicking the band structure of ZnO, and the monolayer 2D material hexagonal boron nitride (\hBN). The LG SBEs are found to require the smooth and periodic structure gauge, while the VG SBEs (without dephasing and separation of current) is independent of the structure gauge. While the VG SBEs require more bands to convergence the HHG spectrum, we find in the case of \hBN{} that the high-order harmonics above the band gap converges faster than the low-order harmonics. Finally, we show that the inclusion of the Berry connections and TDPs in \hBN{} is crucial for the correct treatment of HHG with LG SBEs.

This paper is organized as follows. Section~\ref{sec:theory} contains all the theoretical frameworks pertinent to this work: Sec.~\ref{sec:theory_struc} details the construction of different structure gauges in a general $D$-dimensional material; Secs.~\ref{sec:theory_sbelg} and \ref{sec:theory_sbevg} covers the LG and VG SBEs, respectively; Sec.~\ref{sec:theory_complexity} discusses the computational complexity for the LG and VG SBEs. In Sec.~\ref{sec:zno1d}, the structure gauge construction and the HHG with the two laser gauges are considered for the example of a 1D mathieu model. In Sec.~\ref{sec:hbn2d}, the structure and laser gauges are considered for the more realistic system of monolayer \hBN.
Appendix~\ref{app:1} contains numerical details on the gauge construction.
Atomic units are used throughout this work unless indicated otherwise.

\section{Theoretical methods} \label{sec:theory}

A solution of the SBEs usually proceeds in three steps: (a) solve the time-independent Schr\"odinger equation (TISE) to find the Bloch band energies $E_n^{\vec{k}}$ and Bloch functions  $\phi_n^{\vec{k}}(\vec{r}) = u_n^{\vec{k}}(\vec{r}) e^{i\vec{k}\cdot\vec{r}}$; (b) construct a relevant structure gauge for the Bloch states such that key quantities like the transition dipole and Berry connections can be calculated; (c) solve the SBEs in the desired laser gauge.

For step (a), some form of diagonalization procedure is required. Indeed, many commercial solid state structure codes are highly optimized toward this purpose, e.g. by employing density-functional theory \cite{wien2k, Kresse1996}. In this work, we diagonalize the TISE in reciprocal space
\begin{equation}
  \label{eq:theory_struc_10}
  \sum_{\vec{G'}}\left[ \frac{1}{2} |\vec{k} + \vec{G} |^2 + \mathcal{V}_{\vec{G}-\vec{G}'} \right] u_{n\vec{G}'}^{\vec{k}}
  = E_n^{\vec{k}} u_{n\vec{G}}^{\vec{k}},
\end{equation}
where $\vec{G}$ are the reciprocal lattice vectors, and $\mathcal{V}_{\vec{G}}$ and $u_{n\vec{G}}^{\vec{k}}$ are respectively the Fourier components of a pseudopotential and $u_{n}^{\vec{k}}(\vec{r})$.

Step (b) is detailed in Sec.~\ref{sec:theory_struc}, where the construction of different structure gauges in a general $D$-dimensional crystal is presented, and the calculation of key quantities such as the transition dipoles and Berry connections are discussed. Sections~\ref{sec:theory_sbelg}-\ref{sec:theory_complexity} are about step (c), where the SBEs in the two different gauges are presented and their computational complexities compared.

\subsection{Structure gauges and calculations} \label{sec:theory_struc}

In a solid, the position operator $\vop{r}$ in the Bloch basis can be treated by the identity
\begin{equation}
  \label{eq:theory_struc_1}
  \bra{\phi_m^{\vec{k}}}\vop{r}\ket{\phi_n^{\vec{q}}}_{\text{crys}}
  = \delta_{\vec{k} \vec{q}}\left[ i\delta_{mn}\nabla_{\vec{k}} + \vec{d}_{mn}^{\vec{k}}\right],
\end{equation}
where the subscript ``crys'' refers to integration over the whole crystal, and the dipole matrix elements are given by
\begin{equation}
  \label{eq:theory_struc_1b}
  \vec{d}_{mn}^{\vec{k}} = i \bra{u_m^{\vec{k}}} \nabla_{\vec{k}} \ket{u_n^{\vec{k}}},
\end{equation}
where the inner product $\bra{\cdot}\cdot\ket{\cdot}$ denotes integration over a unit cell. The Berry connections \cite{Xiao2010, Vanderbilt2018} are defined as the diagonal elements \footnote{For those not familiar with the concept of Berry connections, a loose parallel can be drawn to the permanent dipole term in molecular physics, which is also nonzero in systems without inversion symmetry.}
\begin{equation}
  \label{eq:theory_struc_1c}
  \vec{\berryc}_n^{\vec{k}} \equiv \vec{d}_{nn}^{\vec{k}}.
\end{equation}
Due to the gradient in Eq.~\eqref{eq:theory_struc_1b}, to calculate quantities such as the Berry connections $\vec{\berryc}_n^{\vec{k}}$ [Eq.~\eqref{eq:theory_struc_1c}] or propagate the LG SBEs [see Sec.~\ref{sec:theory_sbelg}], the construction of a smooth and periodic structure gauge [see Eq.~\eqref{eq:intro_1}] for the Bloch functions is thus required. We note that many of the published SBE-calculations of solid-state HHG have employed a variety of approximations such as the usage of constant dipole couplings (e.g. \cite{Vampa2014, Vampa2015, Luu2016, Li2019}), neglect of the Berry connections (e.g. \cite{Jiang2017, Jiang2018, Zhang2019}), and neglect of the transition dipole phases altogether. We here illuminate a method for the construction of a smooth and periodic structure gauge that allow the calculations of the Berry connections and TDPs.

For a general $D$-dimensional crystal, we reformulate the $D$-dimensional problem into a series of one-dimensional scalar problems. Suppose that the 1st BZ is spanned by the (generally non-orthogonal) primitive reciprocal lattice vectors $\{\vec{b}_d\}_{d=1}^{D}$ such that all points in the 1st BZ can be written
\begin{equation}
  \label{eq:theory_struc_2}
  \vec{k}=\sum_d^D\kappa_d\vec{\hat{b}}_d, \qquad \kappa_d\in \left[ -\frac{|\vec{b}_d|}{2}, \frac{|\vec{b}_d|}{2} \right], 
\end{equation}
with $\vec{\hat{b}}_d$ the unit vector along $\vec{b}_d$.
Our desired periodic gauge condition reads explicitly
\begin{equation}
  \label{eq:theory_struc_2d}
  \ket{\phi_n^{\vec{k}+\vec{b}_d}} = \ket{\phi_n^{\vec{k}}}.
\end{equation}
Using the expression for the gradient in a nonorthogonal basis, the Berry connections can be written
\begin{equation}
  \label{eq:theory_struc_3}
  \vec{\berryc}_n^{\vec{k}} = \sum_{d_1,d_2=1}^D\berryc_{n,\kappa_{d_1}}^{\vec{k}}g^{d_1 d_2}\vec{\hat{b}}_{d_2},
\end{equation}
with $g^{d_1 d_2}$ the inverse metric tensor, and $\berryc_{n,\kappa_d}^{\vec{k}}$ the scalar Berry connections along the reduced coordinates $\kappa_d$
\begin{equation}
  \label{eq:theory_struc_4}
  \berryc_{n,\kappa_d}^{\vec{k}} = i\bra{u^{\vec{k}}_n}\partial_{\kappa_d}\ket{u^{\vec{k}}_n}.
\end{equation}
We construct a smooth and periodic gauge by adopting the approach discussed in the work of Vanderbilt and co-workers (see Ref.~\cite{Vanderbilt2018} and references therein). This is done in two steps: first a smooth gauge is constructed without periodic boundaries, then using this gauge, the smooth and BZ-periodic gauge is constructed.

Starting from $\ket{u_n^{\vec{k}}}$ obtained from a diagonalization procedure with random phases, we first construct the \textit{parallel transport} (PT) gauge by imposing on the transformed states $\ket{\tilde{u}_n^{\vec{k}}}$ the constraints
\begin{equation}
  \label{eq:theory_struc_5}
  \tilde{\berryc}_{n,\kappa_d}^{\vec{k}} = i \bra{\tilde{u}^{\vec{k}}_n} \partial_{\kappa_d}\ket{\tilde{u}^{\vec{k}}_n} = 0, \qquad \text{for all } d.
\end{equation}
In such a gauge, the adjacent states $\ket{\tilde{u}_n^{\vec{k}}}$ and $\ket{\tilde{u}_n^{\vec{k}+\delta\vec{k}}}$ are ``maximally aligned'', in the sense that their overlap is taken to be strictly real and positive. This alignment is most easily seen from the discrete form of the Berry connection $ \tilde{\berryc}_{n,\kappa_d}^{\vec{k}} \propto \Im\ln\braket{\tilde{u}^{\vec{k}}_n}{\tilde{u}^{\vec{k}+\delta\vec{k}}_n}$ (see Table~\ref{tab:theory_struc_1} and Appendix~\ref{app:1}).

Operator matrix elements such as $\tilde{\vec{d}}_{mn}^{\vec{k}}$ in the PT gauge are smooth inside the BZ, but does not satisfy the periodic gauge condition given by Eq.~\eqref{eq:theory_struc_2d}.
The gauge constraints in Eq.~\eqref{eq:theory_struc_5} are too restrictive, and when wrapping around the BZ boundariies, a Berry phase is accumulated
\begin{equation}
  \label{eq:theory_struc_7}
  \varphi^B_{n,\kappa_d} = \oint_\text{BZ}\tilde{\berryc}_{n,\kappa_d}d\kappa_d.
\end{equation}

The \textit{twisted parallel transport} (TPT) gauge consists of distributing this Berry phase evenly onto the $\ket{\tilde{u}_n^{\vec{k}}}$ obtained from the PT gauge, such that
\begin{equation}
  \label{eq:theory_struc_8}
  \ket{\bar{u}_n^{\vec{k}}} \equiv e^{-i\varphi^B_{n,\kappa_d} \kappa_d / |\vec{b}_d|}\ket{\tilde{u}_n^{\vec{k}}}.
\end{equation}

For 2D and 3D solids, if the TPT gauge is chosen along $\kappa_d$, a subsequent construction of the TPT along another direction $\kappa_b$ with $b \ne d$ would change the gauge along $\kappa_d$, but the resulting gauge would still be periodic along $\kappa_d$.

For the actual discrete numerical procedure to construct the parallel and twisted parallel transport gauges, we refer the reader to Appendix~\ref{app:1}. Here we only stress a crucial point: it is the constructed Bloch functions in the TPT gauge that are BZ-periodic [see Eq.~\eqref{eq:theory_struc_2d}], not the $\ket{u_n^{\vec{k}}}$, such that
\begin{equation}
  \label{eq:theory_struc_9}
  \ket{\bar{u}_n^{\vec{k}+\vec{b}_d}} = e^{-i\vec{b}_d\cdot \vec{r}}\ket{\bar{u}_n^{\vec{k}}},
\end{equation}
which needs to be kept in mind when constructing the gauges and taking the overlaps across BZ boundaries. In Table~\ref{tab:theory_struc_1} we summarize the formulas for the Berry connection and Berry phases in the continuous and discrete cases, as well as in the TP and TPT gauges for the discrete case.

\begin{table*}[t]
  \caption{\label{tab:theory_struc_1}
    Expressions for the Berry connections and phases in the continuous (Column 1) and discrete (Columns 2-4) cases. Columns 3 and 4 list the expressions in the PT and TPT structure gauges. For readability, the real variable $\lambda$ is a placeholder for the $\kappa_d$ variable in the main text, and the crystal momenta $\vec{k}$ and band indices are omitted in the notations for the Berry connections $\berryc_{n,\kappa_d}^{\vec{k}}$ and Berry phases $\varphi_{n,\kappa_d}^B$. It is assumed that $\lambda$ runs over the BZ with length $L$ discretized on an equidistant grid with the last point omitted and spacing $\Delta \lambda$. $N$ is the number of discretization points and $\lambda^j$ is the $j$'th discretization point.}
  \begin{ruledtabular}
    \begin{tabular}{ c | c | c | c}
      continuous & discrete & (discrete) parallel transport & (discrete) twisted parallel transport \\
      \hline
      $\berryc_\lambda = \bra{u_\lambda} i \partial_\lambda \ket{u_\lambda} $
                 & $\berryc_{\lambda^j} = - (\Delta \lambda)^{-1}\text{Im}\ln\braket{u_{\lambda^j}}{u_{\lambda^j+\Delta \lambda}}$
                            & $\tilde{\berryc}_{\lambda^j} = (\Delta \lambda)^{-1} \delta_{j, N-1} \varphi^B $
                                                            & $\bar{\berryc}_{\lambda^j} = \varphi^B L^{-1} $ \\
      \hline
      $\varphi^B = \oint \berryc_{\lambda}d\lambda$
                 & $ \varphi^B = \Delta \lambda \sum_{j=0}^{N} \berryc_{\lambda^j}$
                            & $\varphi^B = \tilde{\berryc}_{\lambda^{N-1}} \Delta \lambda $ 
                                                            & $\varphi^B = \bar{\berryc}_{\lambda^j} L $ \\
    \end{tabular}
  \end{ruledtabular}
\end{table*}

It should be noted that presently we have only constructed the gauges for situations where the considered manifold of energy bands are ``separate'', in the sense that they are nondegenerate over the entire BZ. In the case of degenerate manifolds, we refer the reader to the discussion in Ref.~\cite{Vanderbilt2018}.

\subsection{Semiconductor Bloch equations in length gauge} \label{sec:theory_sbelg}

The SBEs governing a solid driven by a strong laser field in the LG reads \cite{Vampa2014, Floss2018}
\begin{equation}
  \label{eq:theory_sbelg_1}
  \begin{aligned}
    \dot{\rho}_{mn}^{\vec{K}}(t)
    =& -i\left[E_m^{\vec{K}+\vec{A}(t)}-E_n^{\vec{K}+\vec{A}(t)} - \frac{i(1 - \delta_{mn})}{T_2} \right]\rho_{mn}^{\vec{K}}(t)
    \\
    &- i\vec{F}(t)\cdot \sum_l
    \left[\vec{d}_{ml}^{\vec{K}+\vec{A}(t)} \rho_{ln}^{\vec{K}}(t) 
      - \vec{d}_{ln}^{\vec{K}+\vec{A}(t)} \rho_{ml}^{\vec{K}}(t)  \right],
  \end{aligned}
\end{equation}
with $\vec{K}$ the crystal momenta in a reciprocal reference frame moving with the vector potential $\vec{A}(t)$, $\rho_{mn}^{\vec{k}}$ the density matrix elements, $\vec{F}(t) =-\partial_t \vec{A}(t)$ the electric field and $T_2$ the dephasing time. Note that in Eq.~\eqref{eq:theory_sbelg_1}, the Berry connections are included [see Eq.~\eqref{eq:theory_struc_1c}]. We have also tested evaluation of the LG SBEs in the fixed reference frame \cite{Golde2008, Kira2012, Schubert2014} and found that it requires more k-discretization points in order to converge. All results presented in this work is obtained with Eq.~\eqref{eq:theory_sbelg_1}.

The interband and intraband currents read
\begin{equation}
  \label{eq:theory_sbelg_4}
  \begin{aligned}
    \vec{j}_{\text{ter}}(t)
    = & - \int d\vec{K} \sum_{m\ne n} \rho_{nm}^{\vec{K}} \vec{p}_{mn}^{\vec{K}+\vec{A}(t)} \\
    \vec{j}_{\text{tra}}(t)
    = & - \int d\vec{K} \sum_{n} \rho_{nn}^{\vec{K}} \vec{p}_{nn}^{\vec{K}+\vec{A}(t)},
  \end{aligned}
\end{equation}
with $\vec{p}_{mn}^{\vec{k}} = \bra{\phi_m^{\vec{k}}} \hat{p} \ket{\phi_n^{\vec{k}}}$ the momentum matrix elements which are related to the dipole matrix elements in Eq.~\eqref{eq:theory_struc_1b} by
\begin{equation}
  \label{eq:theory_sbevg_2}
  \vec{d}_{mn}^{\vec{k}} = \frac{-i \vec{p}_{mn}^{\vec{k}}}{E_m^{\vec{k}} - E_n^{\vec{k}}}, \qquad \text{for } m\ne n.
\end{equation}
The HHG spectrum is taken as the modulus squares of the Fourier transforms of the currents (after weighting by a window function).

\subsection{Semiconductor Bloch equations in velocity gauge} \label{sec:theory_sbevg}

In the VG, and in the dipole approximation and the absence of electron scattering/correlation, the different $\vec{k}$s are uncoupled and the SBEs read
\begin{equation}
  \label{eq:theory_sbevg_1}
  \begin{aligned}
    \dot{g}_{mn}^{\vec{k}}(t)
    =& -i\left[E_m^{\vec{k}}-E_n^{\vec{k}} \right]g_{mn}^{\vec{k}}(t)
    \\
    &- i \vec{A}(t)\cdot \sum_l
    \left[\vec{p}_{ml}^{\vec{k}} g_{ln}^{\vec{k}}(t)
      - \vec{p}_{ln}^{\vec{k}} g_{ml}^{\vec{k}}(t)  \right],
  \end{aligned}
\end{equation}
with $g_{mn}^{\vec{k}}$ the density matrix elements and
$\vec{p}_{mn}^{\vec{k}} = \bra{\phi_m^{\vec{k}}} \hat{\vec{p}} \ket{\phi_n^{\vec{k}}}$ the momentum matrix elements.
The total current is
\begin{equation}
  \label{eq:theory_sbevg_3}
  \begin{aligned}
    \vec{j}(t)
    = & - \int d\vec{k} \sum_{mn} g_{nm}^{\vec{k}}\left[\vec{p}_{mn}^{\vec{k}} + \delta_{mn}\vec{A}(t) \right].
  \end{aligned}
\end{equation}

\subsubsection{Dephasing and separation of the total current} \label{sec:theory_sbevg_1}

To correctly include the dephasing and separation of the total current into an interband and an intraband contribution, we note that an operator transforms between length and velocity gauges as $\hat{O}^{\LG} = e^{i\vec{A}\cdot\vec{r}}\hat{O}^{\VG}e^{-i\vec{A}\cdot\vec{r}}$ \cite{Foldi2017, Ernotte2018}, with matrix elements
\begin{subequations}
  \label{eq:theory_sbevg_4}
  \begin{align}
    \hat{O}_{mn}^{\LG,\vec{k}+\vec{A}}
    = & \sum_{m'n'} Q_{m'm}^{\vec{k}*} \hat{O}_{m'n'}^{\VG,\vec{k}} Q_{n'n}^{\vec{k}}, \label{eq:theory_sbevg_4a}\\
    Q_{mn}^{\vec{k}}
    \equiv & \braket{u_{m}^{\vec{k}}}{u_{n}^{\vec{k}+\vec{A}}}. \label{eq:theory_sbevg_4b}
  \end{align}
\end{subequations}
In the VG, our procedure to include the dephasing and separation of the current is as follows: for every $n_{\delta t}$ time steps ($n_{\delta t}>1$ for computational efficiency), we transform the density matrix elements to the LG, at which point we apply $\rho_{mn}^{\vec{k}} \rightarrow \rho_{mn}^{\vec{k}}e^{-n_{\delta t}\delta t/T_2}$ for $m\ne n$ and calculate the interband and intraband currents according to Eq.~\eqref{eq:theory_sbelg_4}, whereafter we transform back to the VG. The present procedure to obtain gauge-invariant interband and intraband currents in VG has been previously described in Ref.~\cite{Ernotte2018} in the case of a 1D model solid without dephasing, with a single electron initially at $k=0$, and with $k+A(t)$ never exceeding the 1st BZ boundaries. If the valence band is initially fully populated, $\vec{k}+\vec{A}(t)$ can always exceed the 1st BZ boundaries, and as a consequence, to calculate the overlaps $Q_{mn}^{\vec{k}}$ in \eqref{eq:theory_sbevg_4}, Eq.~\eqref{eq:theory_struc_9} needs to be taken into account, as well as the property $u_{n\vec{G}}^{\vec{k}\pm\vec{b}_d} = u_{n,\vec{G}\pm\vec{b}_d}^{\vec{k}}$.

\subsection{Computational complexities and parameter values} \label{sec:theory_complexity}

We briefly comment on the computational scalings for the time-propagation and current calculations in the LG and VG SBEs.

In the case of the SBEs in LG, for each time step $\delta t$, the binary search, linear interpolations of the relevant energies and matrix elements at $\vec{k} = \vec{K}+\vec{A}(t)$ has the leading order scaling of $O\left[N_{\vec{k}}\log(N_{\vec{k}})N_b^2\right]$, with $N_{\vec{k}}$ the number of $\vec{k}$-discretization points and $N_b$ the number of bands. The calculation of the currents in Eq.~\eqref{eq:theory_sbelg_4} scales as $ O\left(N_{\vec{k}}N_b^2\right)$ and the propagation step (either Runge-Kutta or split-operator) involving $\dot{\rho}_{mn}^{\vec{k}}$ scales as $N_{\vec{k}}N_b^3$. Thus for each time-interval $\Delta t = n_{\delta t}\delta t$ where we calculate the current, the total computational complexity is
\begin{equation}
  \label{eq:theory_scal_1}
  f^{\LG} = O\left[n_{\delta t}N_{\vec{k}}N_b^2\left(\log N_{\vec{k}}+N_b\right)\right].
\end{equation}

\begin{figure}
  \centering
  \includegraphics[width=0.5\textwidth, clip, trim=0 0cm 0 0cm]{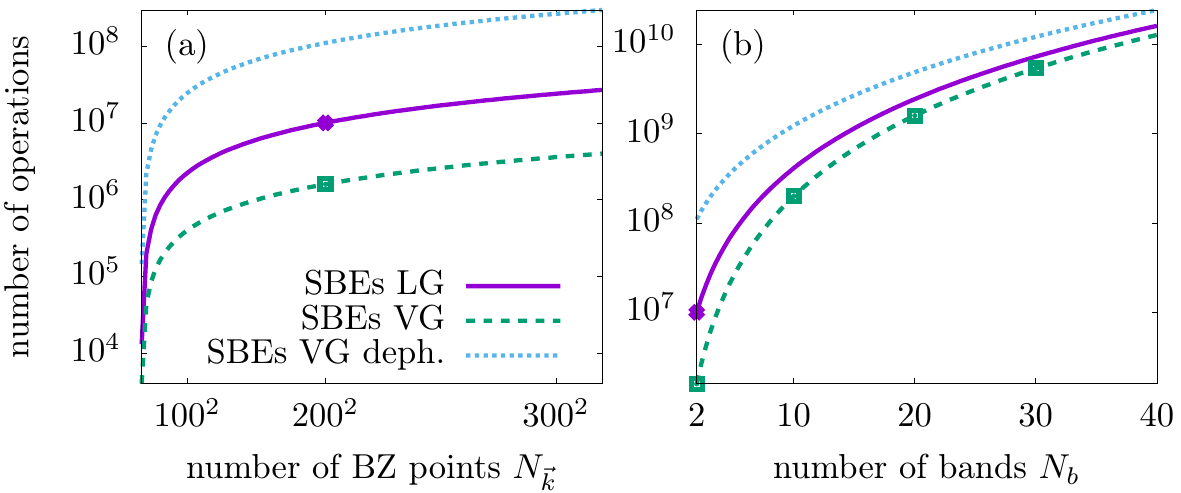}
  \caption{Examples of computational complexities [Eqs.~\eqref{eq:theory_scal_1}-\eqref{eq:theory_scal_3}], with $n_{\delta t}=5$, $N_{\vec{G}}=10$. (a) Fixed $N_b=2$ and varying $N_{\vec{k}}$. (b) Fixed $N_{\vec{k}}=200^2$ and varying $N_b$. Data points are shown for some parameters used for the 2D system in this work (calculations for VG SBE with dephasing is not performed for the 2D system, see text).}
  \label{fig:theory_scal_1}
\end{figure}

In the case of the SBEs in VG without dephasing and separation of the total current, the total computational scaling is
\begin{equation}
  \label{eq:theory_scal_2}
  f^{\VG} = O\left(n_{\delta t}N_{\vec{k}}N_b^3\right).
\end{equation}
When we want to include the dephasing and separate the interband and intraband currents, the complexity increases: the interpolation of the Fourier components $u_{n\vec{G}}^{\vec{k}}$ at $\vec{k}+\vec{A}$ scales as $O\left[ N_{\vec{k}}N_{\vec{G}}N_b\log N_{\vec{k}} \right]$, with $N_{\vec{G}}$ the number of reciprocal lattice vectors; the calculation and storing of the overlaps $Q_{mn}^{\vec{k}}$ in Eq.~\eqref{eq:theory_sbevg_4b} scales as $O\left(N_{\vec{k}}N_{\vec{G}}N_b^2\right)$; the transformation step in Eq.~\eqref{eq:theory_sbevg_4a} scales as $O\left( N_{\vec{k}}N_b^4\right)$; the total complexity is thus
\begin{equation}
  \label{eq:theory_scal_3}
  f^{\VG, \text{deph}} = O\left[ N_{\vec{k}} N_b \left( N_{\vec{G}} + N_bn_{\delta t}\right) \left( \log N_{\vec{k}} + N_b \right) \right].
\end{equation}
It should be mentioned that the complexity formulas are valid in the limit of $N_{\vec{k}}$, $N_{\vec{G}}$, $N_{b}$ going to infinity. For the examples considered in this work, some of these parameter values are quite small, and the formulas only hold approximately. Figure~\ref{fig:theory_scal_1} shows the computational complexities in Eqs.~\eqref{eq:theory_scal_1}-\eqref{eq:theory_scal_3}] plotted for some parameter examples. In our implementation, we parallelize with OPENMP over different $\vec{k}$ values.

For the 1D calculations, we choose $N_{\vec{k}}=600$, $N_b=5$, $N_{\vec{G}}=10$ and $n_{\delta t}=5$ ($\delta t=0.2$), the ratio between complexities of the VG SBEs and the LG SBEs is $f^{\VG}/f^{\LG} = O\left[N_b / (\log N_{\vec{k}} + N_b)\right] = 0.44$. For the case where we want to include the dephasing in VG, the ratio is $f^{\VG,\text{deph}}/f^{\LG} = 1 + N_{\vec{G}} / (n_{\delta t}N_b) = 1.4$.
For the 2D calculations, we use $N_{\vec{k}}=200^2$, $N_b=2$, $N_{\vec{G}}=10^2$, so the ratios are $f^{\VG,\text{deph}}/f^{\LG} = 0.16$ and $f^{\VG}/f^{\LG} = 11$. Note that more states are usually needed in the VG for convergence (but the high-harmonic part of the spectrum could be converged earlier, see Sec.~\ref{sec:hbn2d_hhg}).

\section{Results for 1D ZnO} \label{sec:zno1d}

\begin{figure}
  \centering
  \includegraphics[width=0.5\textwidth, clip, trim=0 0cm 0 0cm]{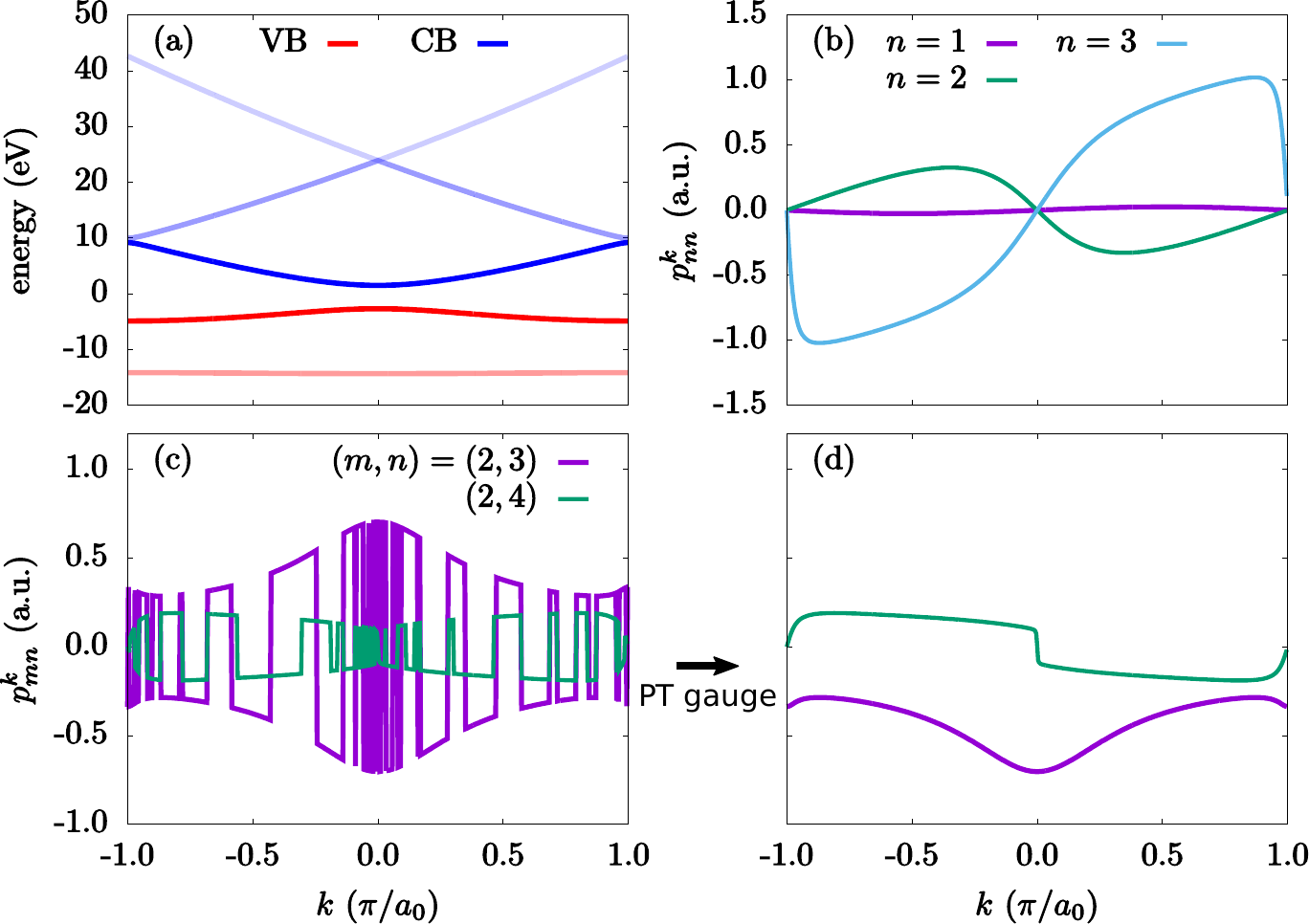}
  \caption{(a) Band structure of the 1D ZnO model. (b) Three diagonal elements of the momentum operator. Two off-diagonal elements of the momentum operator, (c) before and (d) after the application of the parallel transport gauge.}
  \label{fig:zno_struc_1}
\end{figure}

We first consider the 1D ZnO model that has been extensively used in the literature to study HHG in solids, with the Mathieu potential $V(x)=-V_0[1+\cos(2\pi x/a_0)]$, $V_0=0.37$ and $a_0=8$ \cite{Wu2015, Ikemachi2017, LiuXi2017, LiuLu2017, Ernotte2018}. Figures~\ref{fig:zno_struc_1}(a) and \ref{fig:zno_struc_1}(b) show the band structure and a few diagonal elements of the momentum operator $p_{nn}^k$, respectively. Two off-diagonal elements of the momentum operator $p_{mn}^k$, calculated directly after the diagonalization procedure, i.e. without the application of any structure gauge choices, are plotted in Fig.~\ref{fig:zno_struc_1}(c). Since the present 1D potential $\mathcal{V}_G$ is real and symmetric, the $u_{nG}^k$ obtained from Eq.~\eqref{eq:theory_struc_10} are real, resulting in real $p_{mn}^k$. The $\ket{u_{n}^k}$ obtained from the diagonalization procedure assumes a random sign, which is reflected in the discontinuities of $p_{mn}^k$ in Fig.~\ref{fig:zno_struc_1}(c). Construction of the PT gauge from Eq.~\eqref{eq:theory_struc_5} results in the smooth $p_{mn}^k$ shown in Fig.~\ref{fig:zno_struc_1}(d). As we are dealing with real $\ket{u_{n}^k}$, the accumulated Berry phase from Eq.~\eqref{eq:theory_struc_7} is zero, the Berry connections are zero, and the TPT gauge in Eq.~\eqref{eq:theory_struc_8} is the same as the PT gauge.

For the propagation of the SBEs, the lowest two bands in Fig.~\ref{fig:zno_struc_1}(a) are taken as the valence bands and assumed to be initially fully occupied. In the past, many works \cite{Wu2015, Ikemachi2017, LiuXi2017, Ernotte2018} using this model system make the assumption that only electrons with crystal momenta close to $k=0$ contribute to the current, citing the minimum band gap at $k=0$ and the exponential decay of the tunneling probability away from $k=0$. However, recent work (see e.g. Ref.~\cite{Navarette2019}) has shown that other $k$-points contribute significantly to the total HHG yield in terms of the higher conduction bands. In the time propagation, we use the vector potential $A(t) = A_0\cos^2[\pi t/(2\tau)] \cos(\omega_0 t)$, with $A_0=0.30$, carrier frequency $\omega_0=0.0142$ ($\lambda = 3200$ nm) and full width at half maximum (FWHM) time duration $\tau = 48$ fs.

\begin{figure}
  \centering
  \includegraphics[width=0.5\textwidth, clip, trim=0 0cm 0 0cm]{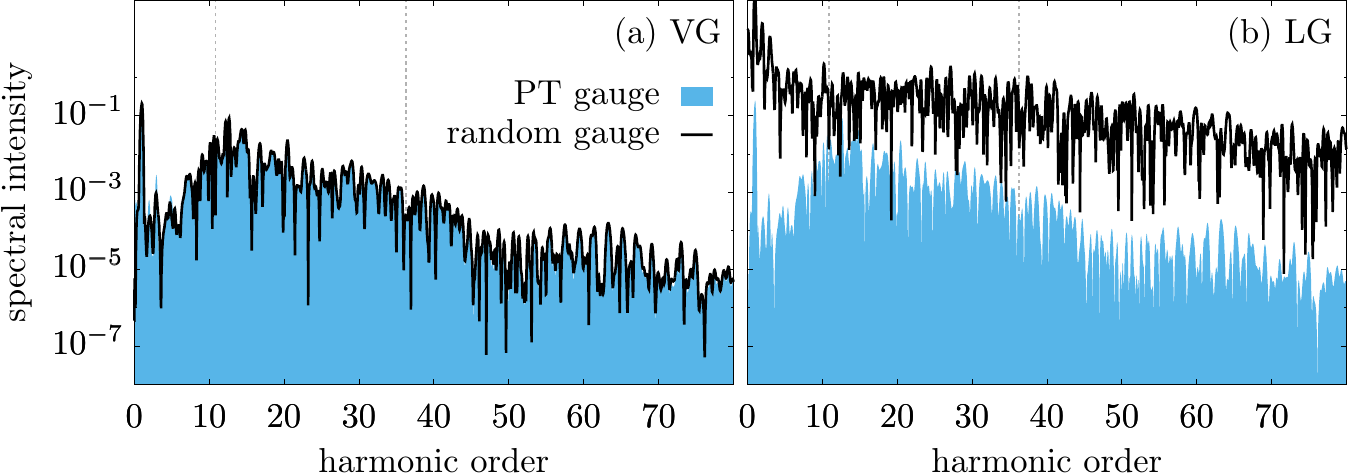}
  \caption{High-harmonic spectrum for the laser-driven Mathieu potential calculated in (a) VG and (b) LG, without dephasing. The vertical dotted lines at around the 11th and 36th harmonics show respectively the band gap energy and the largest energy between the highest valence band and the lowest conduction band. The VG SBEs in (a) is seen to be independent of the structure-gauge choice.}
  \label{fig:zno_hhg_1}
\end{figure}

In the VG SBEs, without the inclusion of dephasing and separation of intraband and interband currents, the HHG spectrum does not depend on the structure gauge choice, see Fig.~\ref{fig:zno_hhg_1}(a). Thus in this specific case (the SBEs for different $k$ are decoupled), the structure gauge construction is unnessesary. The situation is drastically different in the LG SBEs, where the random gauge utterly fails to reproduce the correct spectrum, shown in Fig.~\ref{fig:zno_hhg_1}(b).

\begin{figure}
  \centering
  \includegraphics[width=0.5\textwidth, clip, trim=0 0cm 0 0cm]{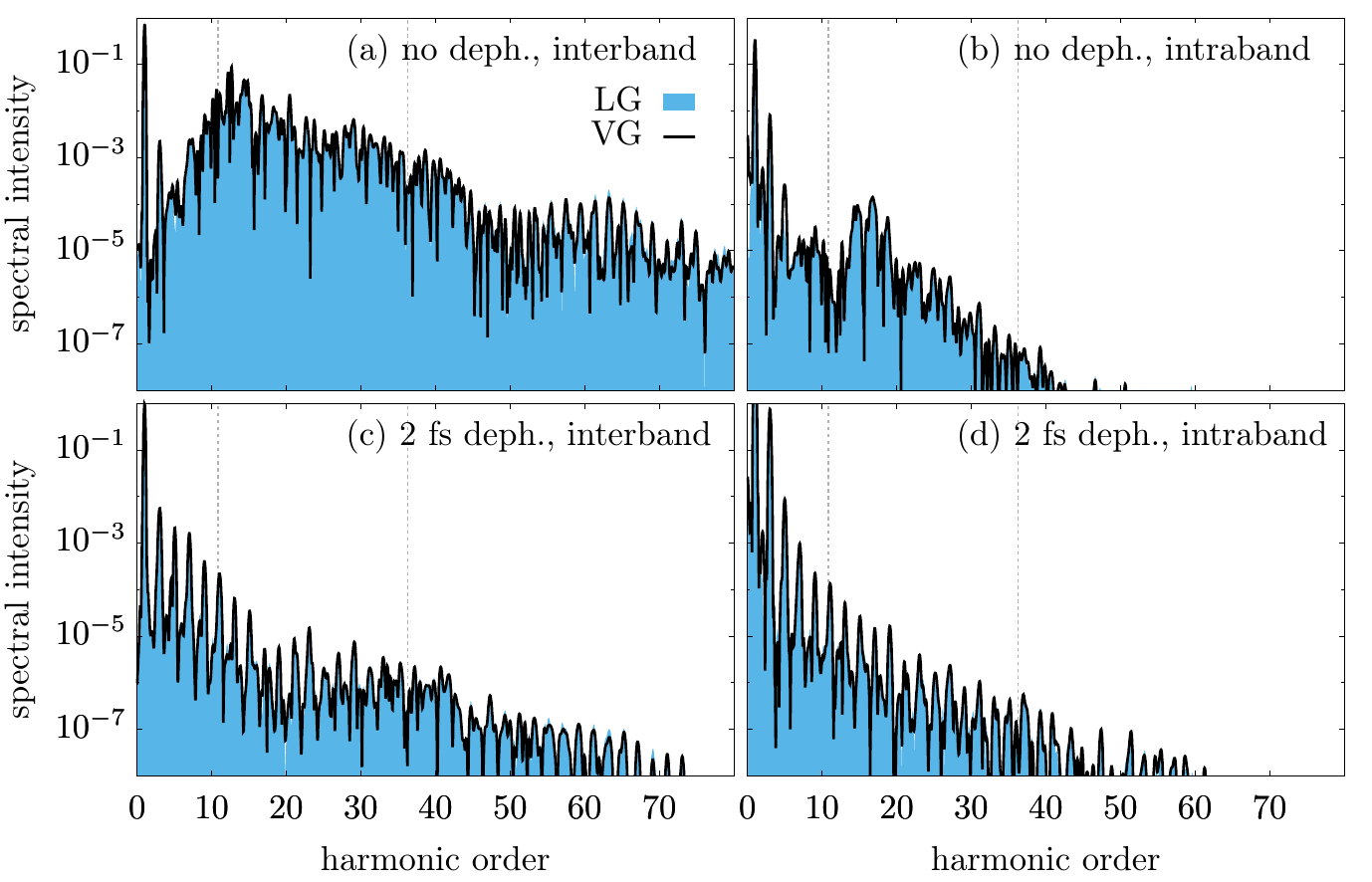}
  \caption{High-harmonic spectrum for the laser-driven Mathieu potential calculated in LG (blue area) and VG (solid line). Upper panels show the (a) interband and (b) intraband contributions for the case of no dephasing. Lower panels (b) and (c) show the result for the case of 2 fs dephasing. The vertical dotted lines at around the 11th and 36th harmonics show respectively the band gap energy and the largest energy between the highest valence band and the lowest conduction band. }
  \label{fig:zno_hhg_2}
\end{figure}

The HHG spectrum for a calculation without the inclusion of dephasing is shown in Figs.~\ref{fig:zno_hhg_2}(a) and \ref{fig:zno_hhg_2}(b) for the interband and intraband contributions, respectively. On the scale of the figure, the LG and VG SBEs agree for harmonics emitted above and below the band gap energy. Without the inclusion of dephasing, destructive interferences lead to significant lower yields in the perturbative region, and almost no well-resolved harmonics are seen in the first plateau region. When a short dephasing time of 2 fs is introduced, we recover the clean harmonics, shown in Figs.~\ref{fig:zno_hhg_2}(c) and \ref{fig:zno_hhg_2}(d). Again, the field gauge results are identical, validating our proposed procedure from Sec.~\ref{sec:theory_sbevg} on the inclusion of dephasing in VG.

\section{Results for monolayer \hBN} \label{sec:hbn2d}

In this section, we consider the case of monolayer \hBN.

\subsection{\hBN{} structure} \label{sec:hbn2d_struc}
\begin{figure}
  \centering
  \includegraphics[width=0.5\textwidth, clip, trim=0 0cm 0 0cm]{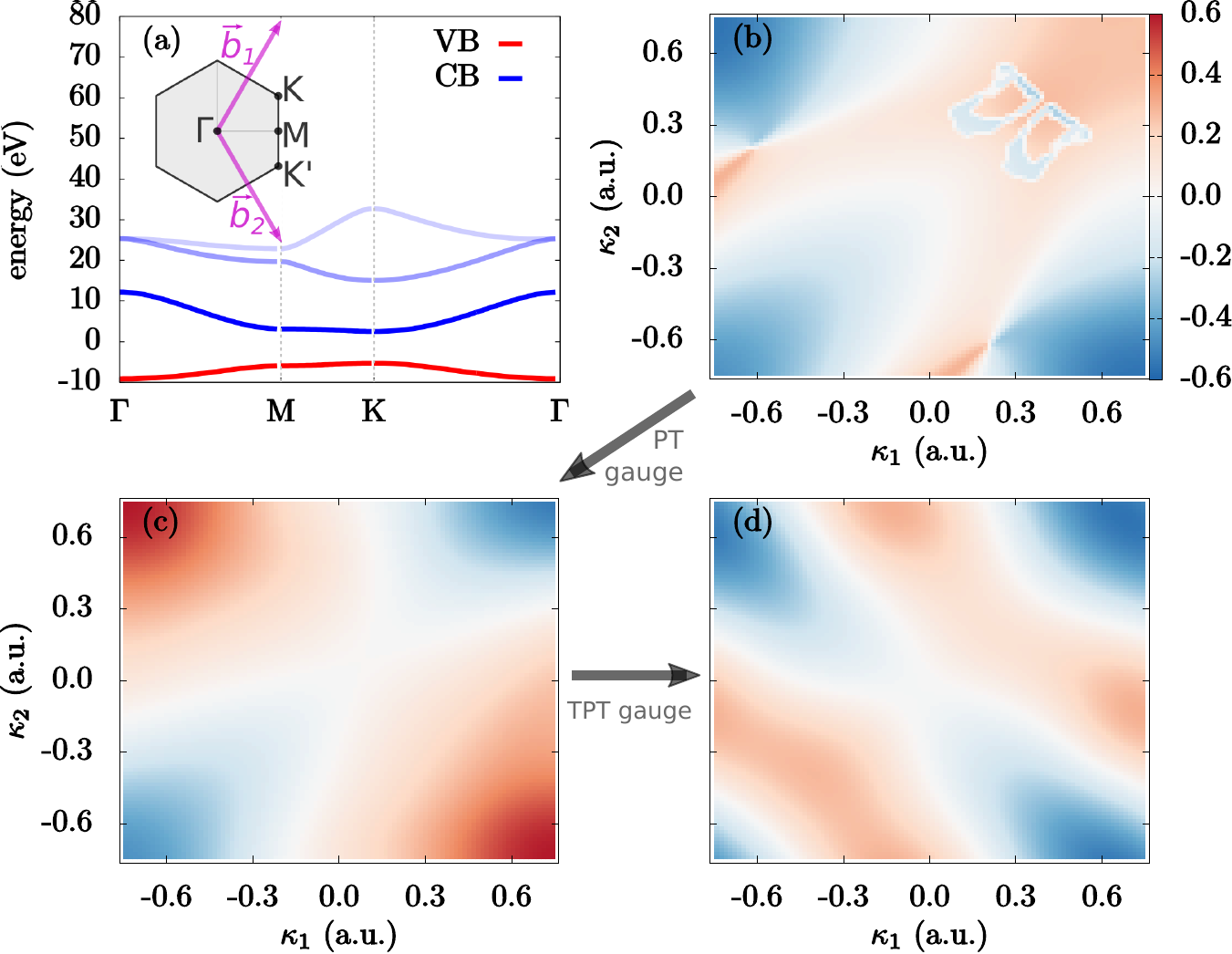}
  \caption{(a) Band structure of the 2D \hBN{} model. (b)-(c) Imaginary part of the $x$-component of the momentum operator, $\text{Im}(\vec{p}_{12, x}^{\vec{k}})$, plotted in the 1st BZ as a function of the reduced coordinates along $\vec{b}_1$ and $\vec{b}_2$ [see Eq.~\eqref{eq:theory_struc_2}]. (b) No structure gauge choice (random gauge); (c) parallel transport gauge; (d) twisted parallel transport gauge.}
  \label{fig:hbn_struc_1}
\end{figure}

For the structure calculation, we employ the pseudopotential method described in \cite{Taghizadeh2017}. Briefly, the Fourier components of the pseudopotential are decomposed into a symmetric and an antisymmetric term,  $\mathcal{V}_{\vec{G}} = \mathcal{V}_{\vec{G}}^S \cos(\vec{G}\cdot \vec{q}) + \mathcal{V}_{\vec{G}}^{AS} \sin(\vec{G}\cdot \vec{q})$, with  $2 \vec{q} = a/\sqrt{3}\uv{e}_x$ and $a=4.7$ the lattice constant.
The {\it form factors} $\mathcal{V}_{\vec{G}}^S$ and $\mathcal{V}_{\vec{G}}^{AS}$ are cut off such that scaled reciprocal lattice vector satisfy $\left|\vec{G}\right|^2 (a/2\pi)^2 \leq 16/3 $. For the four different values $\left|\vec{G}\right|^2  (a/2\pi)^2 =  0$, 4/3, 3, 16/3, the form factors take on the values $\mathcal{V}_{\vec{G}}^S = 0$, 0.39634, 0.05409, 0.28664 and $\mathcal{V}_{\vec{G}}^{AS}=0$, 0.29425, 0, 0.13586, respectively. In the diagonalization of the TISE in Eq.~\eqref{eq:theory_struc_10}, we used 100 reciprocal lattice vectors, which yields a total of 100 bands. In Fig.~\ref{fig:hbn_struc_1}(a), the band structure of the first four bands are shown, with the minimal band gap of $E_b=7.78$ eV located at the K symmetry point.

\begin{figure}
  \centering
  \includegraphics[width=0.5\textwidth, clip, trim=0 0cm 0 0cm]{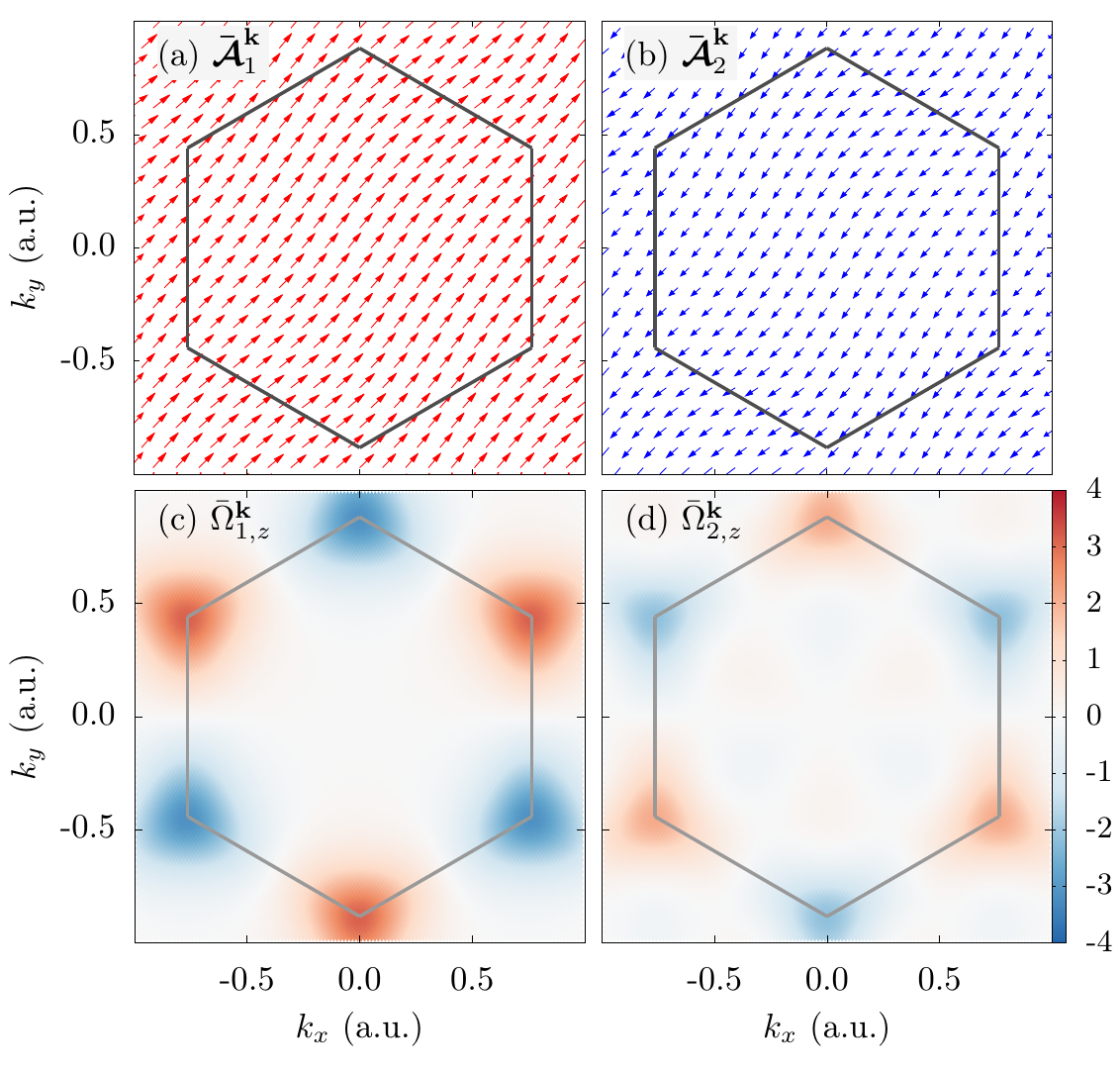}
  \caption{Berry connections for (a) the valence band and (b) the conduction band, in the twisted parallel transport gauge. For better visualization, the arrow sizes in (b) is scaled by two with respect to (a). Berry curvatures for (c) the valence band and (d) the conduction band. The hexagon in the plots guides the eye and traces the 1st BZ.}
  \label{fig:hbn_struc_2}
\end{figure}

To illustrate the properties of the different structure gauges, we consider only the first two bands, taken respectively as the initially fully occupied valence band and the empty conduction band. The PT gauge construction of the higher bands requires the treatment of degenerate states, which is beyond the scope of the present work. In Fig.~\ref{fig:hbn_struc_1}(b)-\ref{fig:hbn_struc_1}(d) the imaginary part of the $x$-component of $\vec{p}_{12}^{\vec{k}}$ are plotted versus crystal momenta $\vec{k}$ in the 1st BZ in the reduced coordinates along $\vec{b}_1$ and $\vec{b}_2$ [Eq.~\eqref{eq:theory_struc_2}]. In the random gauge of Fig.~\ref{fig:hbn_struc_1}(b), there is neither continuity nor BZ-periodicity. The discontinuities are clearly seen around $(\kappa_1, \kappa_2)=(0.3, 0.3)$, while the BZ-non-periodicity is observed when wrapping around the BZ: e.g. along the line $\kappa_2 = 0.6$, where $\text{Im}(\vec{p}_{12, x}^{\vec{k}})$ take on different values at $\kappa_1=-\abs{\vec{b}_1}/2=-0.76$ and $\kappa_1=\abs{\vec{b}_1}/2=0.76$.
In the PT gauge of Fig.~\ref{fig:hbn_struc_2}(c), continuity is ensured without BZ-periodicity, while in the TPT gauge of Fig.~\ref{fig:hbn_struc_2}(d), both continuity and BZ-periodicity are ensured.

The Berry connections in the TPT gauge for the first two bands are plotted in Figs.~\ref{fig:hbn_struc_2}(a) and \ref{fig:hbn_struc_2}(b). The Berry connections are structure-gauge-dependent, and transforms as $\vec{\mathcal{\tilde{A}}}_n^{\vec{k}}=\vec{\berryc}_n^{\vec{k}} - \nabla_{\vec{k}}\varphi_n^{\vec{k}}$ under a general structure-gauge transformation in Eq.~\eqref{eq:intro_1}. The $z$-component of the Berry curvature, $\vec{\Omega}_n^{\vec{k}}\equiv \nabla_{\vec{k}} \times \vec{\berryc}_n^{\vec{k}}$, for the valence and conduction band is shown in Figs.~\ref{fig:hbn_struc_1}(c) and \ref{fig:hbn_struc_1}(d), respectively. They are seen to be prominent around the K symmetry points, and satisfy $\vec{\Omega}_n^{\vec{k}} = -\vec{\Omega}_n^{-\vec{k}}$ due to time-reversal symmetry of our Hamiltonian \cite{Xiao2010, Vanderbilt2018}. Contrary to the Berry connections, the Berry curvatures are structure-gauge invariant, and are responsible for the anomalous velocity which can give rise to a Hall current perpendicular to the applied field polarization direction \cite{Xiao2010}. Recently, extreme nonlinear processes in solids such as HHG and high-order sideband generation have been suggested as a probe to measure the Berry curvature \cite{Banks2017, Luu2018}.

\subsection{\hBN{} HHG} \label{sec:hbn2d_hhg}

\begin{figure}
  \centering
  \includegraphics[width=0.5\textwidth, clip, trim=0 0cm 0 0cm]{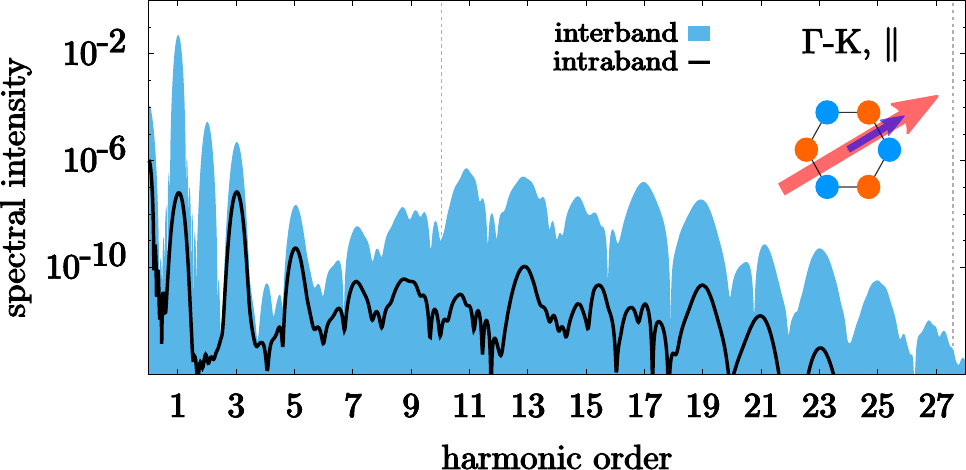}
  \caption{The intraband and interband contribution to the HHG spectrum, for a linearly polarized driver along $\Gamma$-K and parallely polarized harmonics, calculated with LG SBEs.
    The vertical dotted lines at around the 10th and 28th harmonics show respectively the smallest band gap at the K point and the maximal band gap at the $\Gamma$ point. Dephasing time is set to $\infty$. The dominance of the interband contribution over the intraband one is true for all driver and harmonic emission directions considered in this work.}
  \label{fig:hbn_hhg_1}
\end{figure}

\begin{figure}
  \centering
  \includegraphics[width=0.5\textwidth, clip, trim=0 0cm 0 0cm]{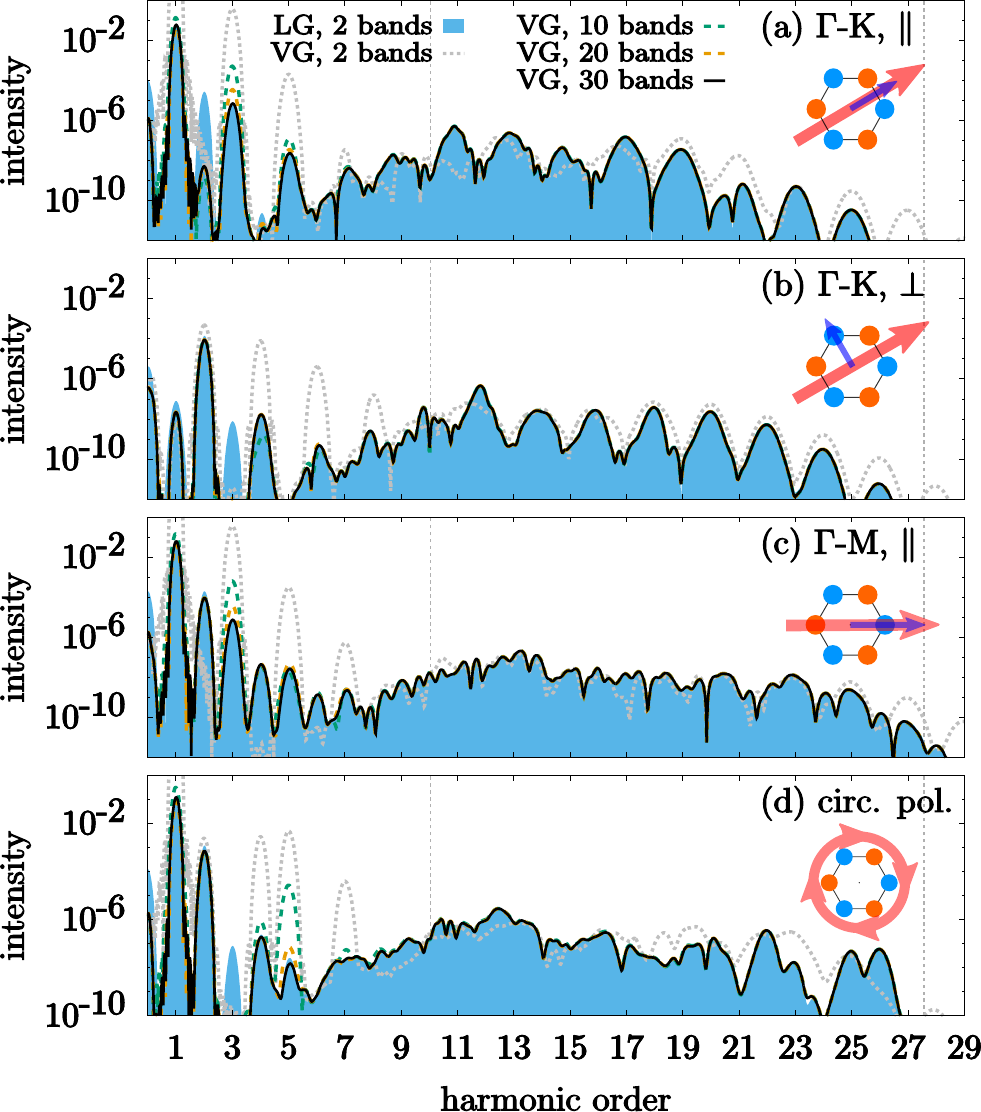}
  \caption{LG and VG convergence properties for HHG in \hBN{} with respect to the number of bands, with (a,b) linearly polarized driver along $\Gamma$-K, (c) along $\Gamma$-M directions, and (d) a circular driver. The insets show the crystal structure, as well as the driver (large red arrows) and emission polarization directions (small blue arrows). The laser parameters are $\lambda = 1600$ nm, $A_0 = 0.35$, $\tau = 29.4$ fs. The vertical dotted lines at around the 10th and 28th harmonics show respectively the smallest band gap at the K point and the maximal band gap at the $\Gamma$ point. Dephasing time is set to $\infty$. Note the fast convergence of the harmonics above the band gap energy compared to below.}
  \label{fig:num_hbn_3}
\end{figure}

For the \hBN{} calculations we use a vector potential of the form
\begin{equation}
  \label{eq:hbn_hhg_1}
  \vec{A}(t) = A_0 g(t) \left[s \uv{e}_{x} \cos(\omega_0 t) + \uv{e}_{2} \sin(\omega_0 t)\right]
\end{equation}
where $A_0=0.35$, $\omega_0=0.0285$ ($\lambda = 1600$ nm), $g(t) = \cos^2\left[\pi t/(2\tau) \right]$ with $t\in[-\tau,\tau]$ and $\tau = 29.4$ fs. For linear polarization $s=0$, while for circular polarization $s=1$ and $\uv{e}_2=\uv{e}_y$. For our current system and pulse parameters used, the interband current strongly dominates over the intraband current for all harmonic orders, see Fig.~\ref{fig:hbn_hhg_1}.

Figure \ref{fig:num_hbn_3} shows the HHG spectra for different driver polarizations, calculated in the \LG{} (with TPT structure gauge) and \VG{} (with random structure gauge). Consider first the \LG{} results given by the filled curves. For the LPD along the $\Gamma$-K direction, purely odd-order (even-order) harmonics are polarized along the parallel (perpendicular) direction and above the band-gap energy, as shown in Fig.~\ref{fig:num_hbn_3}(a) [Fig.~\ref{fig:num_hbn_3}(b)]. When the LPD is along the $\Gamma$-M direction, only high harmonics parallel to the LPD are emitted, which is shown Fig.~\ref{fig:num_hbn_3}(c). For the HHG spectrum using a circular driver in Fig.~\ref{fig:num_hbn_3}(d), above the $\sim20$th harmonic, two pairs of harmonic peaks [(22$\omega_0$, 23$\omega_0$) and (25$\omega_0$, 26$\omega_0$)] are present, with elements in each pair having opposite helicities. This is a consequence of the threefold rotational symmetry of the crystal lattice, and is similar to HHG in atoms and molecules using counterrotating bicircular fields with threefold spatio-temporal symmetry, where the HHG spectrum exihibits combs in which every third harmonic is missing \cite{Fleischer2014, Kfir2015, Medisauskas2015, Baykusheva2016, Mauger2016}. Below the $\sim 20$th harmonic, the harmonic peaks are not well-resolved due to interference of more $\vec{k}$-points in the BZ contributing to a given harmonic energy.

For all the driver and emission polarizations in Fig.~\ref{fig:num_hbn_3}, the convergence properties of the \VG{} SBEs with respect to the number of bands are similar, as we will discuss below. Using only two bands, the shape of the \VG{} high-harmonics deviate considerably from the \LG{} results above the band gap energy. Below the band gap energy, the \VG{} yields are severely overestimated, and in the $\Gamma$-M direction [Fig.~\ref{fig:num_hbn_3}(c)], no even harmonics are generated, in agreement with Ref.~\cite{Taghizadeh2017}. The convergence of the spectrum in \VG{} with respect to the \LG{} is much faster for harmonic orders above the band gap energy: 10 bands are required for convergence above the band gap while 30 bands are required below. This somewhat unintuitive behavior is a manifestation of the different physical mechanism governing the two different harmonic frequency regimes. Above the band gap, the HHG can be explained by the recollision model \cite{Vampa2014, Vampa2015}: an electron tunnels from the valence band into the conduction band, leaving behind a hole; the electron and hole are accelerated by the laser in their respective bands and recollide when they reencounter each other spatially, emitting a photon with energy corresponding to the instantaneous band gap at recollision. Below the band gap, however, the harmonics can be explained by a perturbative analysis \cite{Taghizadeh2017, Dimitrovski2017} involving infinite sums over all band indices, where the \VG{} SBEs is known to diverge in the DC limit. We note that even when using 30 bands in the \VG{} SBEs, there is a mismatch between the \LG{} and \VG{} results below the band gap, which we think could be due to more bands needed in the \LG{} SBEs or numerical inaccuracies in the construction of the relevant gauge-dependent quantities. The fast convergence of the spectrum in the \VG{} above the band gap energy versus below can be exploited in future numerical simulations where one is only interested in the harmonics above the band gap, and does not need to firstly converge the low-order harmonics. Note, however, that our case study of monolayer \hBN{} represents a system where the interband contribution dominates significantly over the intraband one for all harmonic orders. A general convergence statement for cases where the interband and intraband dynamics are competing would need further investigation.

\subsection{Role of the Berry connections and TDPs}

\begin{figure*}
  \centering
  \includegraphics[width=0.75\textwidth, clip, trim=0 0cm 0 0cm]{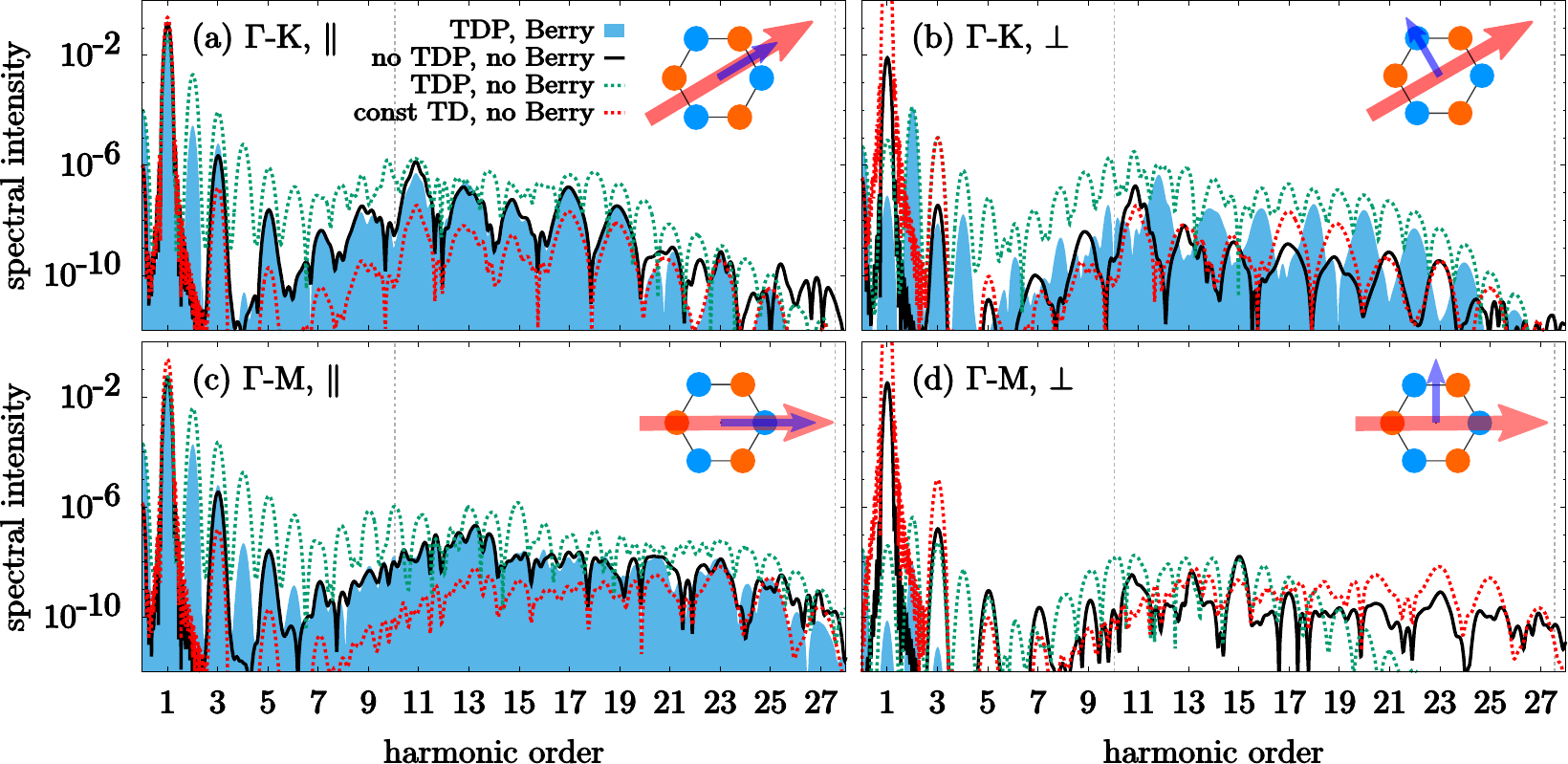}
  \caption{HHG spectra with and without common approximations used in the literature (see text), for \hBN{} driven by linearly polarized pulses. LPD is along (a,b) $\Gamma$-K and (c,d) $\Gamma$-M directions. Laser parameters and figure formats are the same as in Fig.~\ref{fig:num_hbn_3}.}
  \label{fig:hbn_approx_1}
\end{figure*}

We have succesfully demonstrated the construction of a smooth and BZ-periodic gauge, and showed that the two-band LG SBEs correctly produce converged spectra above the band gap. One can now ask the question: why is this important?

In this subsection, we discuss some of the common approximations made in the literature with respect to the Berry connections and TDPs in the \LG{} SBEs and their potential shortcomings. The calculations in this subsection are performed in the \LG{} and TPT gauge, with the dephasing time set to $\infty$ and the pulse parameters identical to those in the previous subsection.

A prevalent approximation in the literature is to neglect both the Berry connections and TDPs,
\begin{subequations}
  \label{eq:hbn_approx_1}
  \begin{align}
    \vec{\mathcal{\tilde{A}}}_n^{\vec{k}} & \equiv \vec{0} \\
    \arg(\vec{\tilde{d}}_{mn}^{\vec{k}}) & \equiv \vec{0}, \quad \text{with} \quad m\ne n.
  \end{align}
\end{subequations}
As mentioned in Secs.~\ref{sec:intro} and \ref{sec:theory_struc}, the use of this approximation is a consequence of the difficulties in constructing a smooth periodic structure gauge. For example, from a basic application of a commercial crystal-structure code, one can obtain the TDPs in the random gauge, and the easiest way to construct a BZ-periodic TDP would be to take the absolute value of the TDPs.

We first apply the approximation in Eq.~\eqref{eq:hbn_approx_1} to monolayer \hBN{} using linearly polarized drivers, with the corresponding HHG spectra plotted in Fig.~\ref{fig:hbn_approx_1} in black solid lines. When the driver is polarized along $\Gamma$-K and $\Gamma$-M, the parallely polarized harmonics shown respectively in Figs.~\ref{fig:hbn_approx_1}(a) and Figs.~\ref{fig:hbn_approx_1}(c) are reproduced quite well overall (compare with the filled curve for the full calculation), with deviations appearing for the highest-order harmonics with orders $\gtrsim 21$. In contrast, the perpendicularly polarized harmonics differ significantly from the full SBEs results: for the $\Gamma$-K driver in Fig.~\ref{fig:hbn_approx_1}(b), purely odd harmonics are emitted instead of the purely even harmonics in the full calculation; while for the $\Gamma$-M driver in Fig.~\ref{fig:hbn_approx_1}(d), nonzero high-harmonic yields are observed even when the full calculation shows no yield. The generation of purely even  or odd harmonics depends on whether the emission of neighbouring half-cycles are exactly in or out of phase \cite{Yue2020arxiv, Jiang2017}. Our results for the perpendicular case in Fig.~\ref{fig:hbn_approx_1}(b) and \ref{fig:hbn_approx_1}(d) show that the omission of the Berry connection and TDPs substantially changes the phase at the time of emission.

In Refs.~\cite{Jiang2017, Jiang2018}, Jiang and coworkers theoretically illustrated the importance of the TDPs for the generation of even-order high harmonics in gapped graphene and wurtzite ZnO, but without inclusion of the Berry connections \footnote{We note that for certain laser-solid symmetry configurations, the Berry connection can be chosen as zero.}. We test the same approximation here, i.e. by using in our simulations the correct TDPs, but setting $\vec{\mathcal{\tilde{A}}}_n^{\vec{k}} \equiv \vec{0}$. The results are shown in Fig.~\ref{fig:hbn_approx_1} by the green dotted lines. For all driver and emission polarization directions, the spectrum contains both odd and even harmonics, and the harmonic yields are severely overestimated. Thus the TDPs and Berry connections should both be included consistently in LG SBE simulations: taking only one of them into account can potentially alter the symmetry properties of the system and result in wrong HHG spectra. Our current results thus disagree with one of the conclusions from Ref.~\cite{Jiang2017}, i.e. the claim that the TDPs can be neglected when the crystal has reflection symmetry with the symmetry plane perpendicular to the LPD.

We also considered another frequently-employed approximation in the literature (see e.g. \cite{Vampa2014, Vampa2015, Luu2016, Li2019}), in which the Berry connections are neglected ($\vec{\mathcal{\tilde{A}}}_n^{\vec{k}} \equiv \vec{0}$) and the transition dipoles are assumed to be constants, assuming the values corresponding to the minimal band-gap crystal momentum (the K symmetry point in our case). This approximation [red dotted line in Fig.~\ref{fig:hbn_approx_1}] produces parallel HHG spectra that have mostly correct shapes but with underestimated yields, and perpendicular HHG spectra that are again quite at odds with the full calculation.

\begin{figure}
  \centering
  \includegraphics[width=0.5\textwidth, clip, trim=0 0cm 0 0cm]{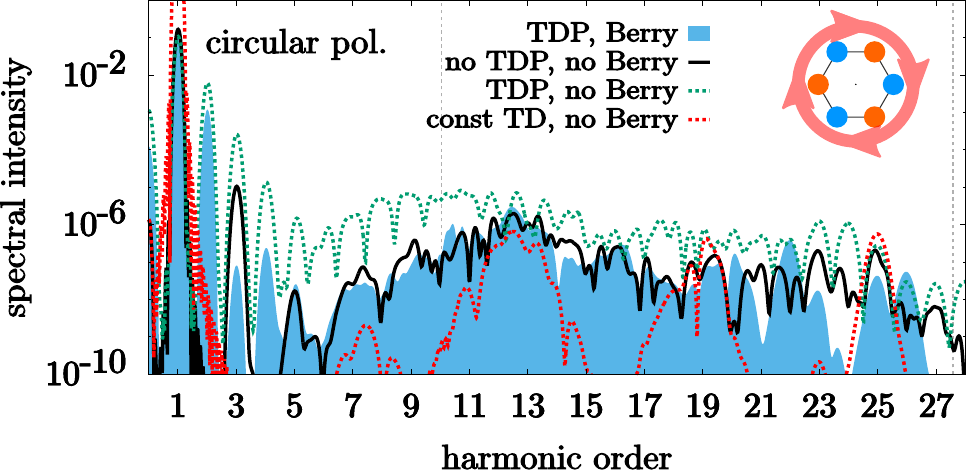}
  \caption{Same format as in Fig.~\ref{fig:hbn_approx_1}, now for a circularly polarized pulse.}
  \label{fig:hbn_approx_2}
\end{figure}

Finally, we calculated the spectra with the above approximations for circularly polarized drivers, which is shown in Fig.~\ref{fig:hbn_approx_2}. In this case, the results for all the three approximations differ significantly from the full calculations. For example, the approximation in Eq.~\eqref{eq:hbn_approx_1} which worked okay in Figs.~\ref{fig:hbn_approx_1}(a) and \ref{fig:hbn_approx_1}(b) fails for the circular polarized case in Fig.~\ref{fig:hbn_approx_2}. Also, the approximation with the constant transition dipole matrix elements exhibits a spectrum involving peaks separated by $6\omega$, wrongly suggesting a crystal structure with six-fold rotational symmetry. Given the recent experimental interest in HHG in solids using circular or elliptical polarized drivers, the usage of such approximations (see e.g. \cite{Zhang2019, Li2019}) should be performed with extreme care.

We have thus shown, at least for the case of \hBN{}, that all the different approximations break down for the perpendicularly polarized harmonics and for circularly polarized drivers. We stress that the applicability of a given approximation highly depends on the considered system. For example, in many bulk solids, tunneling can be expected to occur near the $\Gamma$ symmetry point of the BZ, and the usage of a constant transition dipole is in this case somewhat justified. More investigations are needed for other systems, but we emphasize that rigorous checks and care should always taken when neglecting either the Berry connections or the TPDs.

\section{Summary and outlook}
We have theoretically investigated how the choice of structure and laser gauges affect the SBEs for the HHG process solids. Towards this purpose, we presented a method for the construction of a smooth and BZ-periodic structure gauge, as well as Berry connections which are ubiquitous in systems with time-reversal or inversion symmetry breaking. The LG SBEs was found to be computational more efficient than the VG SBEs, provided that dephasing and separation of the total current are to be included in the calculations. The LG SBEs, however, requires the construction of a smooth and periodic structure gauge. We tested two example systems, a 1D Mathieu model and a pseudopotential model for monolayer hexagonal \hBN. The VG SBEs was found to require more bands for convergence, with faster convergence of the high-order harmonics above the band-gap energy as compared to below. For \hBN{} without inversion symmetry, we showed how the neglect of the Berry connections and TDPs leads to incorrect HHG physics.
We stress that the Berry connections and TDPs are generally nonzero in systems without parity or time-reversal symmetry \cite{Xiao2010, Vanderbilt2018} (e.g. ZnO, \hBN, InP), and their omittance in calculations can potentially lead to qualitatively wrong results. Even in situations when the Berry connections are identically zero, the gauge construction scheme presented in this work is still useful for the construction of smooth matrix elements (e.g. as shown in Sec.~\ref{sec:zno1d}). Our results should be important for all future applications of SBEs for HHG in solids.

\appendix

\section{Numerical details on construction of the PT and TPT gauges} \label{app:1}

In Sec.~\ref{sec:theory_struc}, the construction of the PT and TPT gauges are outlined for the case where $\vec{k}$ is a continuous variable. Here we give more details on the discrete case that can be useful for numerical applications.

Suppose we have discretized the reduced coordinates $\kappa_d$ along $\hat{\vec{b}}_d$ [see Eq.~\eqref{eq:theory_struc_2}] as
\begin{equation}
  \label{eq:app_gauge_1}
  \kappa_d^{j} = \kappa_d^0 + j \Delta\kappa_d, \quad j = 0, \dots, N_d-1
\end{equation}
with $\Delta\kappa_d = \abs{\vec{b}_d} / N_d$ and $N_d$ the number of discretization points. For notational convenience, for now, let the variable $\lambda$ be a placeholder for $\kappa_d$, and the explicit writing of the crystal momenta $\vec{k}$ and band indices be omitted. We have
\begin{equation}
  \label{eq:app_gauge_2}
  \begin{aligned}
    -\Im & \left[ \ln \braket{u_\lambda}{u_{\lambda+\Delta\lambda}} \right] \\
    = & -\Im \left[ \ln \bra{u_\lambda} \Bigl(\ket{u_{\lambda}} + \Delta\lambda \partial_\lambda\ket{u_\lambda} + \cdots \Bigr) \right] \\
    = & -\Im \left[ \ln \left( 1 + \Delta\lambda \bra{u_\lambda} \partial_\lambda \ket{u_\lambda} + \cdots \right) \right] \\
    \approx & - \Delta\lambda  \Im \left[ \bra{u_\lambda} \partial_\lambda \ket{u_\lambda} \right] \\
    = & \Delta\lambda \left[i \bra{u_\lambda} \partial_\lambda \ket{u_\lambda} \right] \\
    = & \Delta\lambda \berryc_{\lambda}, 
  \end{aligned}
\end{equation}
where we have used that $\bra{u_\lambda} \partial_\lambda \ket{u_\lambda}$ is purely imaginary. The Berry connections at the discretazation point $\lambda^j$ [Eq.~\eqref{eq:app_gauge_1}] can thus be evaluated as \cite{Vanderbilt2018}
\begin{equation}
  \label{eq:app_gauge_3}
  \berryc_{\lambda^j} = - (\Delta\lambda)^{-1} \Im \left[ \ln \braket{u_{\lambda^j}}{u_{\lambda^j+\Delta\lambda}} \right].
\end{equation}

The continuous and discrete cases, for the PT and TPT gauges, are summarized in Table.~\ref{tab:theory_struc_1} of the main text.

\acknowledgements

The authors acknowledge support from the National Science Foundation, under Grant No. PHY1713671.  L. Y. thanks Francois Mauger for useful discussions.



%

\end{document}